# Gravity Balanced Arm Exoskeleton for Basketball Shooting Training

Yunfei Liu and Zhanghao Yang

*Abstract*— This paper proposes a gravity balanced arm exoskeleton design for basketball shooting training. The potential energy equation of the mechanism is derived. A simulation of the arm going through the basketball shooting motion is done on the mechanism. Throughout the motion the total potential energy is constant. Thus, the proposed arm exoskeleton is indeed gravity balanced with the use of two springs.

## I. INTRODUCTION

According to [1], over 26 million people play basketball in the United States. Being able to shoot a basketball is an essential skill for players to score throughout the game. One can improve their basketball shooting percentage by increasing the number of shots they take, repetitions, and having proper technique [2,3]. Combining Iterative Learning Control (ILC) [4,5,6] with basketball shooting practice involves using feedback from each shot to continuously refine technique. Sensors capture shot data, and an ILC algorithm adjusts the player's technique based on errors from previous shots. This personalized, data-driven approach improves shooting accuracy and efficiency over time.

There is a lack of documentation on devices that can assist in basketball shooting training. One of the few devices that was found was a patent for a training device [7], but there was no research study done on the device.

This paper proposes a gravity balanced arm exoskeleton to assist players in basketball shooting training. To be gravity balanced, the device must have a constant potential energy regardless of the configuration [8]. There are several designs that have used springs to obtain this gravity balance [9,10], however the designed proposed varies in the fact that there are several links but only two springs needed to gravity balance the mechanism.

The goal for the arm exoskeleton is to assist players in basketball shooting practice. With the arm exoskeleton supporting its own mass and the mass of the player's arm, the player can ideally do more shooting repetitions since they will not tire as easily. The gravity balance arm exoskeleton will also assist the players in having the proper shooting technique, as described in [3].

## II. GRAVITY BALANCE OF A BASKETBALL TRAINER

### A. Design

The arm exoskeleton is designed to support the human arm while a player shoots a basketball. To support the arm, the mechanism must maintain gravity balance during the entire motion. The design of the mechanism is a planar solution because the ideal shooting technique requires the player's arm to remain aligned with their shoulder while shooting a basketball [3]. The design is a modified two link planar mechanism with two springs, Fig. 1.

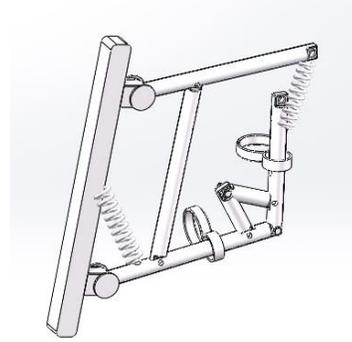

Figure 1. Gravity Balanced Arm Exoskeleton CAD Model

### B. Calculations and Assumptions

To derive the potential energy equations the following assumptions were made:

- All joints are revolute.
- Ignore frictional forces at the revolute joints.
- Ignore the mass of springs.
- Assume the natural length of springs are zero.
- Assume the center of gravity is at the geometric center of each link.
- Assume the arm of the player increases the mass of link 1 and link 2.

The variables in Table I are used to derive the potential energy equation.

TABLE I

| Symbol | Definition |
| --- | --- |
| $g$ | Gravitational Acceleration |
| $m_i$ | Mass of the Links (i=1,2,3,4,5,6) |
| $l_i$ | Length of the Links (i=1,2,3,4,5,6) |
| $q_i$ | Angle of Links (i=1,2) |
| $k_i$ | Stiffness Coefficient of Springs (i=1,2) |
| $b_i$ | Distance between the fixed point of spring one and the distance from the first revolute joint (i=1,2) |
| $l_s$ | Distance along link 2, the length of where spring two is connected |

| $l_t$ | Distance along link 6, the length to where link 5's revolute joint is attached |

As shown in Fig. 2, this gravity balanced system has six links and two springs. The design of this system only introduces two angle variables, (q1 and q2), thus it is a 2-degree of freedom mechanism, equivalent to a 2-link mechanism. The reason why the design has link 3 and link 4, is to increase the moving stability of the system, (the gradient of gravitational potential energy is smoother). Link 5 and link 6 were introduced to balance the mass of link 2 and the forearm of a player.

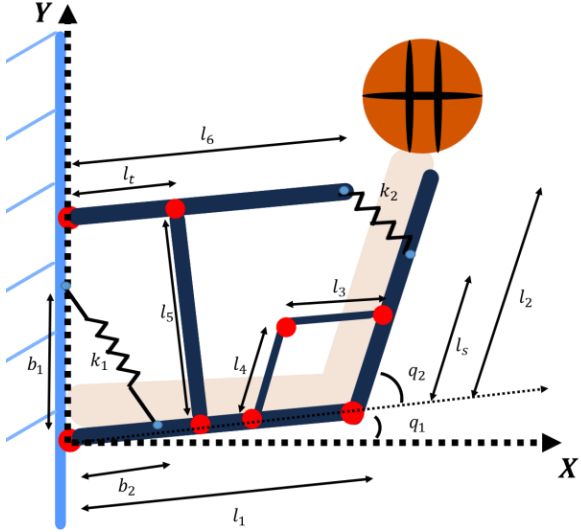

Figure 2. Gravity Balanced mechanism with links and springs labeled

To simplify the derivations of the potential energy equation the following relations, (1-6), were assumed about the architecture of the gravity balance mechanism.

$$l_2 = 0.9 l_1 \tag{1}$$

$$l_1 = l_5 = l_6 \tag{2}$$

$$l_3 = l_4 = \frac{1}{3} l_1 \tag{3}$$

$$b_1 = 2 b_2 = \frac{4}{9} l_1 \tag{4}$$

$$l_t = \frac{1}{3} l_1 \tag{5}$$

$$l_s = \frac{35}{36} l_2 \tag{6}$$

Then the gravitational potential energy (7) and elastic potential energy (8) were calculated as:

$$V_G = g(m_1 \frac{l_1}{2} \sin q_1 + \\
m_2 \left(l_1 \sin q_1 + \frac{l_2}{2} \sin(q_1 + q_2)\right) + \\
m_3 l_1 \left(\frac{5}{6} \sin q_1 + \frac{1}{3} \sin(q_1 + q_2)\right) + \\
m_4 l_1 \left(\frac{2}{3} \sin q_1 + \frac{1}{6} \sin(q_1 + q_2)\right) + \\
m_5 \left(\frac{l_1}{3} \sin q_1 + \frac{l_5}{2}\right) + m_6 \left(\frac{l_1}{2} \sin q_1 + l_5\right)) \tag{7}$$

$$V_S = \frac{1}{2} k_1 l_1^2 \left(\frac{20}{81} - \frac{16}{81} \sin q_1\right) + \frac{1}{2} k_2 (l_5^2 + l_s^2 - 2 l_5 l_s \sin(q_1 + q_2)) \tag{8}$$

To enable the system to be balanced, the coefficient of sinq1 and sin(q1+q2) should satisfy the following conditions in (9) and (10):

$$l_1 g \left(\frac{1}{2} m_1 + m_2 + \frac{5}{6} m_3 + \frac{2}{3} m_4 + \frac{1}{3} m_5 + \frac{1}{2} m_6\right) = \frac{8}{81} k_1 l_1^2 \tag{9}$$

$$m_2 g \frac{l_2}{2} + m_3 g \frac{l_1}{3} + m_4 g \frac{l_1}{6} = k_2 l_5 l_s \tag{10}$$

Thus, the total potential energy is a constant, which means the mechanism is gravity balanced, and the final equation becomes (11):

$$V = V_G + V_S \\
= g l_5 (\frac{1}{2} m_5 + m_6) + \frac{10}{81} k_1 l_1^2 + \frac{1}{2} k_2 (l_5^2 + l_s^2) \tag{11}$$

The total potential energy is thus only a function of the spring constants, mass of the links, and the length of the links.

In order to make this mechanism fit more flexible situations, in the future the design will be considered into a 3-D movement and also equipped with mass springs and frictional joints. To simplify the structure of design and improve the capacity of its moving functions, the number of links and springs will change and counterweights, pulleys, mass strings and prismatic joints will also be reasonably added.

### III. SIMULATION

To verify the design is gravity balanced in all configurations, equations (7-10) were simulated in MATLAB.

The mass of link 1 was simulated as 4.6 kg because it was assumed that it would also bear the weight of the arm. The average weight for total arm is 3.6 kg based on the mean collected from [11]. The mass of link 2 was simulated to have 1 kg.

The length of links 1 and 6 are the same at 0.30 m, which is the mean upper arm length for adults based on [11]. The length of link 2, is 0.27 m which is also the average of human forearms.

With the corresponding masses and lengths, the spring constants were solved for using (9,10). The mechanism was then simulated to move q1 from -π/2 to π/2 and q2 from 0 to π. This simulates the range of motion a basketball player would move their arm, starting at their side, the arm would move upward, while the elbow would bend 90 degrees and back to 0 as the player shoots the basketball. The simulated range of motion depicts that the gravity balance arm exoskeleton can maintain constant total potential energy through each configuration Fig 3.

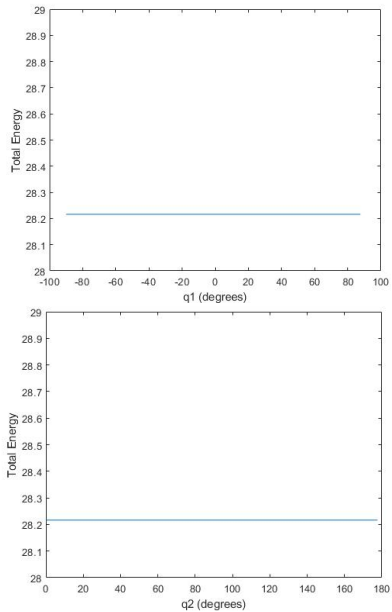

Figure 3. The total potential energy of the mechanism as angle q1 and q2 changes

Figure 4 shows how the gravity and spring potential energy components change as the arm changes its q1 and q2 angles. However, the total potential energy remains constant throughout the changing angles.

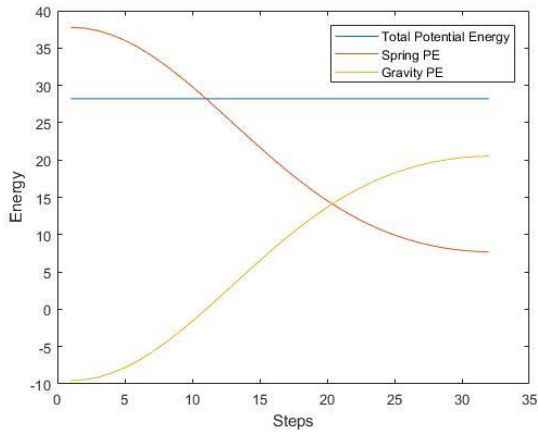

Figure 4. The spring and gravity potential energy as the mechanism moves from all its possible configurations.

The previous simulations were done using the average human arm mass and length. Since basketball players vary in mass and length the spring constants were recalculated for a varying mass of the arm. The mass of the arm was added to link 1 and link 2. Figure 5 shows the relationship between the mass of the links and the spring constant required to keep the mechanism gravity balanced.

Y. Liu is a master's Graduate Student in The Fu Foundation School of Engineering and Applied Science, Mechanical Engineering, Columbia University, New York, NY 10027 USA.

Z. Yang is a master's graduate student in The Fu Foundation School of Engineering and Applied Science, Mechanical Engineering, Columbia University, New York, NY 10027 USA (email: zy2313@columbia.edu).

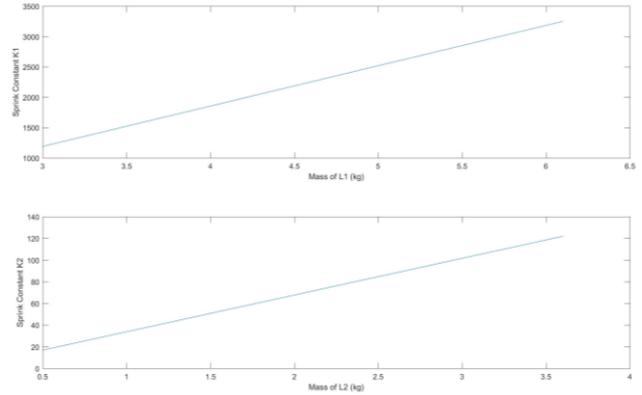

Figure 5. Spring constant 1 and 2 in relations to mass of link 1 and 2

IV. CONCLUSION

This paper proposed a gravity balanced arm exoskeleton that can assist in basketball shooting training. Simulations validated the design's ability to support the mass of the mechanism itself and the mass of the player's arm in all configurations. This can assist players in alleviating their arm's weight, thus one would ideally be able to perform more shooting practice repetitions while maintaining proper shooting technique. In future work, this design can be manufactured and tested on healthy human subjects while they practice their basketball shooting.